\documentclass[reprint,secnumarabic,amssymb, nobibnotes, aps, prl]{revtex4-1}

\usepackage{graphicx}
\usepackage{dcolumn}
\usepackage{bm}
\usepackage{amssymb}
\usepackage{amsmath,amsthm,amsfonts}

\def\R{\mathbb{R}}

\newcommand{\ket}[1]{|\kern.3ex#1\kern.3ex\rangle}
\newcommand{\bra}[1]{\langle\kern.3ex #1 \kern.3ex|}
\newcommand{\scalar}[2]{\langle\kern.3ex #1 \kern.3ex|\kern.3ex#2\kern.3ex\rangle}

\begin{document}

\title{Infrared behavior and gauge artifacts in de Sitter spacetime: The photon field}

\author{A. Youssef}
\email{ahmed.youssef@apc.univ-paris7.fr}
\affiliation{Universit\'e Paris Diderot Paris 7, Laboratoire APC, B\^atiment Condorcet, Case 7020, 75205 Paris Cedex 13, France}
\date{\today}
\begin{abstract}
We study the infrared (long-distance) behavior of the free photon field in de Sitter spacetime. Using a two-parameter family of gauge-fixing terms, we show that the infrared (IR) behavior of the two-point function is highly gauge-dependent and ranges from vanishing to growing. This situation is in disagreement with its counterpart  in flat spacetime, where the two-point function vanishes in the IR region for any choice of the gauge-fixing parameters. A criterion to isolate the ``physical'' part of the two-point function is given and  is shown to lead to a well-behaved two-point function in the IR region. 
\end{abstract}
\maketitle
\paragraph{\textbf{Introduction.}}
The study of quantum field theory in the de Sitter background is of paramount importance to the understanding of  the early Universe as well as its present accelerated expansion. One of the most striking, yet poorly understood,  aspects of de Sitter quantum field theory (QFT) is the so-called \emph{infrared problem}. The simplest and most serious manifestation of this problem being the nonvanishing of the correlation functions for largely - spacelike and timelike -separated points. This is for instance a well-known fact for the massless minimally coupled (mmc) scalar and the graviton fields. In this Letter we study the IR behavior of the photon field in de Sitter space and show that it exhibits similar IR pathologies. More importantly we are able to show that these are purely gauge artifacts. 

The organization of the Letter is as follows. First we compute the correlation function using the $R_\lambda$ gauges $\lambda (\nabla_a A^a)^2/2$. Then we generalize the computation to the larger family of higher derivative gauge-fixing terms $ \lambda (\nabla_a A^a)^2/2 +\xi R^2 (\nabla_a A^a)\Box (\nabla_a A^a)/2$. Finally we propose a decomposition of the two-point function into a physical and a gauge part and we show that the physical part is vanishing in the IR region. 

The $d$-dimensional de Sitter spacetime $X_d$ of ``radius'' $R$ can be identified with the real one-sheeted hyperboloid in the $d+1$ Minkowski spacetime $M_{d+1}$
$$
X_d=\left\lbrace x \in \R^{d+1}, \eta_{\mu \nu}x^\mu x^\nu=R^2\right\rbrace
$$
where $\eta_{\mu \nu}$ is  the ``mostly plus'' flat metric  $\textrm{diag}(-1,+1,\cdots,+1)$. Let $\mu(x,x')$ denote the geodesic distance between the points $x$ and $x'$ and $g_{ab}$ be the d-dimensional de Sitter (or d-dimensional Minkowski if $R\to \infty$) metric. 
We will compute the Wightman two-point function $W_{aa'}(x,x')=\langle\Omega\vert A(x)A(x')\vert \Omega \rangle$ in different gauges. In the previous formula $\vert \Omega \rangle$ designates the Euclidean (Bunch-Davies) vacuum state. This vacuum being de Sitter invariant, $W_{aa'}(x,x')$ is a maximally symmetric bi-tensor. As shown in detail in \cite{Allen86}, all such bi-tensors can be expressed in terms of the parallel propagator $g_{aa'}(x,x')$ between $x$ and  $x'$, and the two vectors $n_a=\nabla_a \mu$ and $n_{a'}=\nabla_{a'} \mu$. Our two-point function is thus written as
$$
W_{aa'}(x,x')=\alpha_1(\mu) g_{aa'}+\alpha_2(\mu) n_a n_{a'} 
$$
where, for spacelike separations ($\mu^2>0$),  $\alpha_1$ and $\alpha_2$ are scalar functions of the geodesic distance $\mu$ only. We will frequently use the following formulas \cite{Allen86}:
\begin{align*}
g_a^{\:\:b'} n_{b'}&=-n_{a'},  \qquad  \qquad   \qquad \:g_{ab}=g_a^{\:\:c'}g_{c' b}\\
\nabla_a n_b&=A (g_{ab}-n_a n_b), \quad \nabla_a n_{b'}=C (g_{ab'}+n_a n_{b'}) \\
 \nabla_a g_{bc'}&=-(A+C) (g_{ab}n_{c'}+g_{ac'}n_b)
 \end{align*}
where 
$$
 A=\frac{\cot(\mu/R)}{R},  \quad \text{and} \quad C=-\frac{\csc(\mu/R)}{R}.
$$
It is useful to introduce the quantity $z$ given for spacelike separations ($\mu^2>0$ or $0<z<1$) by
\begin{equation}
\label{defz}
z=\cos^2 \left( \frac{\mu}{2R}\right).
\end{equation}
All of our calculations are performed in the spacelike region $0<z<1$. They can be extended further by analytic  continuation in the $z$ variable to the cut complex plane $\mathbb{C} \setminus (1,\infty)$. Finally, the Feynman propagator $G^F_{a a'}=i \langle \Omega \vert T A_a(x) A_{a'}(x')\vert \Omega \rangle $ can be obtained as the limiting value of $W_{aa'}$ when approaching the branch cut $z>1$ from above.

We end our introduction by noting that several, quite interesting, recent works (see \cite{MarolfHollands} among many others) were devoted to the study of infrared effects in de Sitter. In these articles however, only massive (although \emph{interacting}) fields are considered. IR pathologies are of course much stronger for massless fields and they appear already at the linear level.
\paragraph{\textbf{The $R_\lambda$ gauges}.}
The gauge fixed action describing the massless vector field $A_a$ is
\begin{equation}
\label{action1}
S_\lambda=\int d^4x \sqrt{-g} \left( \frac{1}{4} F_{ab}F^{ab}+ \frac{\lambda}{2} (\nabla_a A^a)^2\right)
\end{equation}
where $F_{ab}=\nabla_a A_b-\nabla_b A_a$. The resulting equation of motion is $D_{a b} A^b=0$ where 
\begin{equation}
\label{operatorD}
D_{a b}= g_{ab} \Box-\frac{(d-1)}{R^2} g_{ab}+(\lambda-1) \nabla_a \nabla_b.
\end{equation}
\paragraph{$1$. Flat space.}
In flat space ($R \to \infty$), the Fourier transform of the Feynman propagator reads 
$$
G^F_{aa'}(k)=\frac{1}{k^2-i\epsilon} g_{aa'}+\frac{1-\lambda}{\lambda} \frac{1}{(k^2-i\epsilon)^2}k_a k_{a'}.
$$
Using the (massless limit) of the scalar Feynman propagator (given, as always, for spacelike separations) 
\begin{equation}
\label{scalargreen}
G^F(x,x')=\int \frac{d^4k}{(2\pi)^4} \frac{e^{i k (x-x')}}{k^2+m^2-i \epsilon}=\frac{i m}{(2\pi)^2} \frac{K_1(m \mu)}{\mu}
\end{equation}
(where $K$ is the modified Bessel function of the second kind) and its derivatives with respect to $x$ and $x'$, we get the coordinate-space two-point function
\begin{equation}
W_{aa'}= \left[\frac{\lambda +1}{8 \pi ^2 \lambda  \mu ^2}\right] g_{aa'}-\left[\frac{\lambda -1}{4 \pi ^2 \lambda  \mu ^2}\right] n_{a} n_{a'}.
\end{equation}
For all values of $\lambda$, this expression vanishes for large timelike and spacelike separations as expected.

\paragraph{$2$. De Sitter space.}
Following \cite{Allen86}, we write the two-point function in de Sitter space as 
\begin{equation}
\label{generalW}
W^{[\lambda]}_{a a'}(x,x')= \alpha_1^{[\lambda]}(z) \: g_{a a'}+\alpha_2^{[\lambda]}(z) \: n_a n_{a'}. 
\end{equation}
where $\alpha_1,\alpha_2$ are scalar functions of the invariant quantity $z$ defined in \eqref{defz}. The equation of motion $D_{ab} W^{b}_{\:\:a'}=0$ implies two independent equations on $\alpha_1$ and $\alpha_2$. These equations are most easily solved in the Feynman gauge $\lambda=1$ and the result is found in \cite{Allen86} to be
\begin{align*}
\alpha_1^{[\lambda=1]}(z)&=\frac{1}{48 \pi^2 R^2} \left[ \frac{3}{1-z} +\frac{1}{z}
+\left( \frac{2}{z}+\frac{1}{z^2}\right) \ln(1-z)\right]\\
\alpha_2^{[\lambda=1]}(z)&=\frac{1}{24 \pi^2 R^2 } \left[ 1-\frac{1}{z} +\left( \frac{1}{z}-\frac{1}{z^2} \right)\ln(1-z)\right]
\end{align*}
Instead of vanishing for largely separated points, the correlation goes to a nonzero constant function
\begin{equation}
\label{IR1}
\lim_{z\to \infty} W^{[\lambda=1]}_{aa'}(x,x')=\frac{1}{24 \pi^2 R^2 } \: n_a n_{a'}.
\end{equation}
This constant vanishes in the flat spacetime limit ($R\to \infty$) as expected. The nonvanishing of the correlator  in de Sitter space seems to contradict our experience with the more familiar flat space QFT. It is indeed a general result of the latter, sometimes referred to as the cluster decomposition principle \cite{Strocchi}, that the correlation functions must decay in the IR region. To our knowledge, only very special QFT models, e.g., confinement models like the massless scalar in two-dimensional flat space, seem to escape this result.

As explained in the introduction, it turns out that this IR pathological behavior of the photon field is typical of de Sitter massless fields. The most notable examples are the mmc scalar and the graviton fields. The correlation functions of these fields exhibit even stronger pathologies as they grow in the IR region (see, for instance, \cite{Moschella,Staro79,Antoniadis07}). 

This ``IR problem'' is at the heart of an open discussion in the community. It led some authors \cite{Antoniadis07} to speak of a  ``striking violation of cluster decomposition properties of the de Sitter invariant vacuum state.'' It has also been shown that this growing of the graviton two-point function renders scattering amplitudes divergent \cite{ilio87}. As a consequence, the authors of \cite{Antoniadis07}, among many others, suspect  a \emph{quantum instability} of de Sitter space, meaning that the de Sitter geometry is not a stable ground state of quantum gravity with a cosmological constant. 

Other workers in the field, mainly motivated by the fact that the correlations of \emph{some} gauge invariant quantities fall off in the IR limit, claim that the nonvanishing of the two-point functions in the IR region is nothing but a gauge artifact. For instance, the electromagnetic field correlation function is well behaved in the IR region\cite{Allen86}:
$$
\langle F_{ab} F^{a' b'}\rangle = \frac{1}{8 \pi^2 R^4} \frac{1}{(1-z)^2}  \left( g_{[a}^{\:\:[a'}g_{b]}^{\:\:b']} + 4\: n_{[a} g_{b]}^{\:\:[b'} n^{a']} \right).
$$
The same is true for the graviton field as there are no growing terms in the correlation
function of the Riemann tensor. 

In the following we will study in detail the IR problem in the case of the photon field in de Sitter. The IR behavior of this field received considerably less attention in the literature than its counterpart for the mmc and the graviton field. This is certainly due to the importance of these two fields in inflation theory. It is also probably due to the fact that in actual QED calculations the nonvanishing term \eqref{IR1} will not contribute to scattering processes (see the last section).  This model will however prove to be precious in realizing that - at least for some situations - serious IR pathologies in de Sitter space are, without doubt, gauge artifacts.

The equation of motion \eqref{operatorD} applied to the two-point function \eqref{generalW} leads after some algebra to a coupled system of differential equations. Hereafter we will only need one of these equations, namely,
\begin{align}
\label{eom1}
\nonumber &\mathbf{H}\left[1,d-2,\frac{1}{2}+\frac{1-\lambda}{2}\right] \alpha_1+\frac{\lambda -1}{2}  \alpha_2' \\
&+\frac{\lambda -\lambda _d}{\lambda_d-1} \left[ \frac{\alpha _1 }{z}+ \frac{1}{2}\left( \frac{1}{ 1-z}-\frac{1}{ z}\right)\alpha_2\right]=0 
\end{align}
where the prime denotes derivation with respect to the variable $z$, $\lambda_d=\dfrac{d-3}{d-1}$ and  $\mathbf{H}$ is the hypergeometric operator defined by
$$
\mathbf{H}[a,b,c]\alpha= z(1-z)\alpha''+\alpha'+ab \:\alpha
$$
Another, moreover first order, relation between $\alpha_1$ and $\alpha_2$ is obtained by considering the two-point function of the field strength
$$
\langle F_{ab} F^{a' b'}\rangle = \beta_1 \: g_{[a}^{\:\:[a'}g_{b]}^{\:\:b']} + \beta_2 \: n_{[a} g_{b]}^{\:\:[b'} n^{a']}.
$$
In fact, the equation of motion (in vacuum)  $\nabla^a F_{ab}=0$ implies the relation \cite{Allen86}
\begin{align}
\label{relation}
\alpha_2&=\frac{z}{8 \pi ^2 R^2 (z-1)}+2 (1-z) \left(z \alpha_1'+\alpha _1\right)
\end{align}
We now specialize our results to the $d=4$ case for simplicity. The general solution in the Euclidean vacuum is found to be  
\begin{align*}
\label{solution1}
\alpha_1^{[\lambda]}(z)=\frac{1}{48 \pi^2 R^2 \lambda} &\left[ \frac{3(\lambda+1)}{2(1-z)} +\frac{3\lambda-1}{2z}+\right.\\
 & \left.(3\lambda-1)\left( \frac{1}{z}+\frac{1}{2 z^2}\right) \ln(1-z)\right]\\
\alpha_2^{[\lambda]}(z)=\frac{1}{24 \pi^2 R^2 \lambda} &\left[ 1-\frac{3(\lambda-1)}{2(1-z)} -\frac{3\lambda-1}{2z}+\right.\\
&\left.(3\lambda-1)\left( \frac{1}{2z}-\frac{1}{2 z^2} \right)\ln(1-z)\right]
\end{align*}
and the IR behavior is given by
\begin{equation}
\lim_{z\to \pm \infty} W^{[\lambda]}_{aa'}(x,x')=\frac{1}{24 \pi^2 R^2 \lambda} \: n_a n_{a'}.
\end{equation}
This result means that one can cure any IR bad behavior by going to the \emph{Landau gauge} $\lambda \to \infty$. More importantly, we see that the IR behavior of the correlation function is \emph{pure gauge}, thus asking whether or not the two-point function is well or ill behaved in the IR region is in itself -at least partially-a misleading question.  

Finally we note that there exists a special choice of the gauge-fixing parameter, namely $\lambda=\frac{1}{3}$ (more generally $\lambda_d$ in $d$ dimensions), that cancels logarithmic terms and gives a particularly simple two-point function. We note, perhaps as a curiosity for the time being, that for large dimensions $d$, $\lambda_d$ goes to the Feynman gauge $\lambda=1$, which is the simplest gauge in flat space.

\paragraph{\textbf{More pathological gauges.}}
We now consider a larger family of gauge-fixing terms. Our motivation is that the IR pathology  we exhibited in the last section, namely that the correlation function tends to a constant, might seem -I believe mistakenly- mild enough not to worry about. We will show that in this more general gauge, the two-point function is actually growing in the IR region, exactly like the mmc scalar or the graviton fields. 
 We consider the action 
\begin{align}
\label{action2}
S_{\lambda,\xi}=S_\lambda+\frac{\xi}{2 m^2} \int d^4x \sqrt{-g}  (\nabla_a A^a)\Box (\nabla_a A^a)
\end{align}
where we have added a higher derivative gauge-fixing term. Similar higher derivative gauge fixing terms are used in QCD, the electroweak theories, and were studied in \cite{bartoli}. They were also used to study flat space perturbative quantum gravity \cite{ilio86}. We note that if we want $\lambda$ and $\xi$ to be dimensionless, such higher derivative gauge-fixing terms require the introduction of a mass quantity $m$ . In flat space $m$ is introduced by hand, while in de Sitter space the inverse of the de Sitter radius naturally plays this role:  $m^2=\frac{1}{R^2}$. This observation renders the introduction of a higher derivative gauge-fixing somehow more natural in the de Sitter case. Finally note that the use of a two-parameter family of gauge-fixing terms that are simultaneously nonzero does not overconstrain the system's dynamics. In fact, in the Faddeev-Popov procedure, the gauge-fixing  term added to the action is merely the probability distribution $P[\omega]$ of the function $\omega(x)$ that intervenes in the generalized Lorentz gauge condition $\nabla^a A_a=\omega(x)$. This is of course a valid gauge condition and does not overconstrain the dynamics.

The equation of motion reads
\begin{equation}
\label{eom2}
\left[\Box g_{ab}+(\lambda-1) \nabla_a \nabla_b +\frac{\xi}{m^2} \Box  \nabla_a \nabla_b \right]A^b=0
\end{equation}
\paragraph{$1$. Flat space.}
The flat space Feynman propagator is found in Fourier space to be
 \begin{align*}
\tilde{G}^F_{a a'}(k)=&\frac{ g_{aa'}}{k^2-i\epsilon}+\Bigg[\frac{1-\lambda }{\lambda\:  (k^2-i\epsilon)^2 } \\
&+\frac{\xi}{m^2 \lambda ^2} \left( \frac{1}{k^2-i\epsilon}- \frac{1}{k^2-\frac{m^2 \lambda}{\xi}-i\epsilon}\right)\Bigg] k_a k_{a'}
\end{align*}
The reader will note the relative negative sign  typical of higher derivative theories in front of the last propagator.
Using \eqref{scalargreen} and its derivatives the coordinate-space correlator is found in terms of Bessel functions. For all values of the gauge-fixing parameters it vanishes in the IR as expected.
\paragraph{$2$. De Sitter space.}
We now consider the action \eqref{action2} in de Sitter space and thus set $1/m^2=R^2$.
Following the same methods described above we obtain the following equations on  $\alpha_1$:
\begin{align*}
\alpha_1^{(4)}&=\frac{4 (\tau -2) (\tau +3)}{(z-1)^2 z^2} \alpha_1+\frac{3 (2 z-1) \left(\tau ^2+\tau -16\right)}{(z-1)^2 z^2} \alpha_1'\\
&-\frac{12-(z-1) z (\tau -8) (\tau +9)}{(z-1)^2 z^2}\alpha_1''-8 \left(\frac{1}{z}+\frac{1}{z-1}\right) \alpha_1'''\\
+&\frac{\left(\tau ^2+\tau -2\right) (2 (1-3 \lambda ) z-3)}{32 \pi ^2
   \lambda  R^2 (z-1)^4 z^2}-\frac{18-5 \tau  (\tau +1)}{32 \pi ^2
   R^2 (z-1)^4 z^2}
\end{align*}
where we introduced the parameter $\tau=\frac{\sqrt{4 \lambda +9 \xi }-\sqrt{\xi }}{2 \sqrt{\xi }}$. Obtaining this equation requires some elaborate algebraic manipulations. The calculations in this section were thus verified with the tensor algebra system xAct on Mathematica \cite{xAct}. 
This equation can be solved in closed form. Using the asymptotic formulas of Legendre functions, the solution verifying (i) regularity at $z=0$ and (ii) flat space short-distance singularity is found.  For noninteger values of $\tau$ it reads
\begin{align*}
\alpha_1(z)&=\frac{ 2 Q_{\tau }^2(2 z-1)-\pi \cot (\pi  \tau ) P_{\tau }^2(2 z-1)}{32 \pi^2  R^2 \lambda  \:(z-1) z  \left(\tau ^2+\tau -2\right)} \\
   &+\frac{3 \lambda -1}{96 \pi ^2 R^2 \lambda } \left[ \frac{(2 z+1) \ln (1-z)}{z^2}+\frac{1}{z}\right]\\
   &+\frac{1}{32 \pi ^2 R^2 \lambda  (\tau -1) (\tau +2)}\\
   & \left[ \frac{1}{(z-1)^2}-\frac{(\lambda +1) \tau \left(\tau+1 \right)-2 \lambda}{z-1}\right]
     \end{align*}
where $P$ and $Q$ are Legendre functions of the first and second kind, respectively.  As before, $\alpha_2$ is obtained by \eqref{relation}. Using the asymptotic behavior of the  Legendre functions we find for instance that for $\tau >2$ the two-point function grows like $z^{\tau-1}$ in the IR region. For $(\lambda=0,\xi\neq 0)$ the two-point function can be expressed in terms of polylogarithms and the IR behavior is given by 
$$
W_{aa'} \sim -\frac{\ln z }{72 \pi ^2 \xi  R^2 } n_a n_{a'}.
$$
We thus find that it grows logarithmically in the infrared region like the two-point function of the mmc scalar field. This IR growing is purely a gauge artifact since, as proven in the beginning of this Letter, no IR pathologies arise in the Landau gauge. That such a growing behavior is nothing but a gauge artifact for the photon field suggests that the situation might be the same for the mmc and the graviton fields. This possibility will be examined in detail in \cite{Ahmed}.

\paragraph{\textbf{A physical decomposition.}}
\label{sectiondecomposition}
The previously described situation, namely, that the IR behavior of the two-point function in de Sitter space contains important gauge artifacts, makes it natural to look for a decomposition into  physical and nonphysical parts. We show now that such a decomposition exists and is given by the following rewriting of the two-point function:
\begin{equation}
\label{vecphysdecomp}
W^{[\lambda]}_{a a'}(x,x')= g_{aa'}\beta_1(z)+\nabla_a \nabla_{a'} \beta_2(z).
\end{equation}
In fact, most of the physics is included in the ``quantum'' action $W_J$ defined by
$$
\exp\left[\frac{i}{\hbar} W_J\right]=\int D[A_a]  \exp \left[\frac{i}{\hbar} \left(S+\int d^4x \sqrt{-g} \: A_a J^a\right)\right]
$$  
where $J_a$ is an external  conserved current $\nabla_a J^a=0$. The tree-level expression  is given by
$$
W_J=\int dV_x dV_{x'}  J^a(x) W_{a a'}(x,x') J^{a'}(x')
$$
where $dV_x=d^4x \sqrt{-g(x)}$ is the invariant volume element. Since the current $J_a$ is conserved, integration by parts ensures that only $\beta_1$ will contribute to $W_J$ and it thus fully deserves to be referred to as the ``physical'' two-point function.
An explicit form of the functions $\beta_1$ and $\beta_2$ can be found by solving the system
\begin{align*}
\alpha_1=\beta_1+\frac{1}{2R^2} \beta_2', \quad R^2 \alpha_2=(1-z)\beta_2' +z(1-z)\beta_2''.
\end{align*}
The solutions are unique if one requires regularity near $z=0$ and we obtain
\begin{equation}
W^{\textrm{Phys}}_{a a'}(x,x')=\frac{1}{16 \pi^2 R^2} \left(\frac{1}{1-z}+\frac{\ln(1-z)}{z}\right) g_{a a'}
\end{equation} 
It is a remarkable fact that $W^{\textrm{Phys}}_{a a'}$ is $(\lambda,\xi)$ independent and is well behaved in the IR region.
In flat spacetime, the Feynman gauge $(\lambda=1,\xi=0)$ gives precisely this physical two-point function. The situation in de Sitter is quite different and no choice of the gauge-fixing parameters gives the physical two-point function. A natural question is whether a more general gauge-fixing term gives directly $W^{\textrm{Phys}}_{a a'}(x,x')$ in de Sitter space. 
\\
\begin{acknowledgments}
It is a pleasure to thank  E. Mottola, A.M. Polyakov, and A. Roura for their comments on an earlier version of this Letter.
\end{acknowledgments}

\end{document}